\documentclass[12pt,fleqn]{article}
\usepackage{amsmath,amssymb,graphicx,ifthen}
\title{Compact polymers on decorated square lattices}
\author{
Saburo Higuchi\thanks{e-mail: hig@rice.c.u-tokyo.ac.jp}\\
{\it Department of Pure and Applied Sciences,} \\
{\it The University of Tokyo} \\
{\it  Komaba, Meguro, Tokyo 153-8902, Japan}
}

\date{November 30, 1998, revised in March 1999, %
          UT-KOMABA/98-25, cond-mat/9811426}
\renewcommand{\thefootnote}{\fnsymbol{footnote}}
\newcommand{\version}{condmat}

\begin{document}
\begin{titlepage}
\thispagestyle{empty}
  \begin{flushright}
          November 1998\\
(revised) March 1999\\
          UT-KOMABA/98-25\\
          cond-mat/9811426
  \end{flushright}
\vspace*{\fill}

\begin{center}
{\Huge\bf
Compact polymers\\[8pt] on decorated square lattices
}
\vspace*{\fill}

{\Large Saburo Higuchi}%
\footnote{e-mail: \texttt{hig@rice.c.u-tokyo.ac.jp}}\\[0.5ex]

{\it Department of Pure and Applied Sciences,} \\
{\it The University of Tokyo} \\
{\it  Komaba, Meguro, Tokyo 153-8902, Japan}
  \end{center}
\vspace*{\fill}

\vspace*{\fill}
\begin{center}
{\large\bf Abstract}
\end{center}
\smallskip 

A Hamiltonian cycle of a graph is a closed path that visits every
vertex once and only once. It serves as a model of a compact polymer
on a lattice.  I study the number of Hamiltonian cycles, or
equivalently the entropy of a compact polymer, on various lattices
that are not homogeneous but with a sublattice structure.
Estimates for the number are obtained by two methods.  One is the
saddle point approximation for a field theoretic representation.  The
other is the numerical diagonalization of the transfer matrix of a
fully packed loop model in the zero fugacity limit.  In the latter
method, several scaling exponents are also obtained. \vspace*{5ex}

\noindent
PACS:
05.20.-y, 
02.10.Eb, 
05.50.+q, 
82.35.+t 

\noindent
Keywords: 
Hamiltonian cycle; 
Hamiltonian walk;
self-avoiding walk; 
compact polymer;
fully packed loop model
\end{titlepage}

\setcounter{footnote}{0}
\renewcommand{\thefootnote}{\arabic{footnote}}

\section{Introduction}
A Hamiltonian cycle of a graph is a closed self-avoiding walk
which visits every vertex once and only once.
The number of all the Hamiltonian cycles of a graph $G$ is denoted 
by $\mathcal{H}(G)$.
Hamiltonian cycles have often been used to
model polymers which fill the lattice completely \cite{ClJa:polymerE}.
The quantity $\log\mathcal{H}(G)$ 
corresponds to the entropy of a polymer system
on $G$ in the compact  phase. 
One can enrich the model
by introducing a weight that  depends on the shape of cycles
in order to model  more  realistic polymers.
For example, the polymer melting problem is studied by 
taking into account the bending energy \cite{BaGaOr:polymermelting}. 
Another extended model relevant to the protein folding problem 
is proposed in refs. \cite{ChDi:protein,CaTh:minimum,DoGaOr:proteinfolding}.
The quantity $\mathcal{H}(G)$ is also related to the partition
function of zero-temperature O($n$) model in the limit $n\rightarrow0$.

For homogeneous graphs (lattices) with $N$ vertices,
one expects that $\mathcal{H}(G)$ behaves as
\begin{equation}
  \mathcal{H}(G) \rightarrow C(G) N^{\gamma-1} \omega^N 
\quad\quad (N\rightarrow\infty),
\label{asymptotic}
\end{equation}
where the entropy per vertex $\log\omega$ is defined by
\begin{equation}
  \log \omega=\lim_{N\rightarrow\infty} \frac{1}{N}\log \mathcal{H}(G).
  \label{def_of_omega}
\end{equation}
The connectivity constant $\omega$ is supposed to be a bulk quantity
which does not depend on boundary conditions, 
whereas the conformational exponent $\gamma$ depends on such details of
graphs \cite{ScHiKl:compact}.
I assume that eq.\eqref{asymptotic} is not multiplied by a surface
term $\mu^{(N^\sigma)}, \sigma<1$ proposed in
refs.\cite{OwPrBr:newscaling,Malakis:hamiltonian,DuDa:manhattan}
because in the present case
Hamiltonian cycles  fill the whole lattice which itself does not have
a boundary. 

A field theory representation of $\mathcal{H}(G)$
for an arbitrary graph $G$ is introduced 
by Orland, Itzykson, and de Dominicis \cite{OrItDo:hamiltonian} and 
has been used to
study the extended models \cite{BaGaOr:polymermelting,DoGaOr:proteinfolding}
as well as the original one\cite{Suzuki:regular,cond-mat/9711152}.
For homogeneous graphs with the coordination number $q$, 
the saddle point approximation yields
\begin{equation}
  \mathcal{ H}(G) \sim  \left(\omega_{\mathrm{SP}}\right)^N, 
\quad
\omega_{\mathrm{SP}}:=\frac{q}{e}
\label{mean_field_saddle_point_estimate}
\end{equation}
or $\omega\simeq \omega_{\mathrm{SP}}$.
Eq. \eqref{mean_field_saddle_point_estimate} has been very
successful.
For the square lattice, $\omega_{\mathrm{SP}}=4/e=1.4715\cdots$ is
quite near to a numerical estimate $\omega\simeq 1.473\cdots$ 
\cite{ScHiKl:compact,BaBlNiYu:packedloop,DuSa:exact,JaKo:fieldtheory}.
It is found that the estimate $\omega_{\mathrm{SP}}=4/e$ is unaltered
in the next leading order in the perturbation theory \cite{cond-mat/9711152}.
For the triangular lattice, the field theory predicts
$\omega_{\mathrm{SP}}=6/e=2.20728\cdots$ while a numerical calculation suggests
$\omega\simeq2.095\cdots$ \cite{BaBlNiYu:packedloop}.
An exact solution is available for the hexagonal lattice 
\cite{Suzuki:regular,BaSuYu:honyeycomb} . 
It implies $\omega=3^{3/4}/2=1.13975\cdots$, which is near to the
estimate $\omega_{\mathrm{SP}}=3/e=1.10364\cdots$.
More examples are given in ref. \cite{Suzuki:regular}.
The saddle point equation is so good that one may speculate that 
the system is dominated by the saddle point configuration.
It is desirable  to understand why this approximation works so well.

In this article, I systematically study $\mathcal{H}(G)$ for  
\emph{inhomogeneous} graphs by the field theory.
A graph is said to be \emph{homogeneous} if its automorphism group
acts on the set of vertices transitively. Roughly, a graph (lattice) is
homogeneous if all the vertices are equivalent.
Note that the estimate \eqref{mean_field_saddle_point_estimate} is not
applicable to inhomogeneous graphs.
An example of inhomogeneous graph is the square-diagonal lattice shown in
Fig.~\ref{fig:square_diagonal}. 
An earlier study in this direction can be found in ref. \cite{Suzuki:regular}.

Inhomogeneous graphs are as important as homogeneous ones in physics
because they model certain realistic materials well.  
Applying the saddle point method to them is expected to cast light on the 
nature of this extraordinary good approximation.

Another achievement in this work is an accurate measurement of the
connectivity constants $\omega$ and scaling exponents by the numerical
transfer matrix method. This is based on the mapping onto the fully
packed loop model in the zero fugacity limit.  The results should be
compared with the field theoretic estimates.

The organization of the paper is as follows.
In section \ref{sec:field_theory_rep}, the field theoretic
representation of $\mathcal{H}(G)$ is explained in order to make the
presentation self-contained. 
I derive a formula for an estimate of $\mathcal{H}(G)$ by the saddle
point method when a graph $G$ has a sublattice structure which makes
$G$ inhomogeneous in section \ref{sec:sublattice}.
It is applied for a number of examples in section \ref{sec:examples}.
In section \ref{sec:exact_relations}, I prove some exact relations among
$\mathcal{H}(G)$ for several lattices. 
I explain the method and the result of numerical transfer matrix
analysis in section \ref{sec:transfer_matrix}.
I summarize  my results in section \ref{sec:summary}.

\section{Field theoretic representation}  \label{sec:field_theory_rep}
Let $G=(V,E)$ be a graph, where $V$ and $E$ stand for the sets of
vertices and edges.
One sets $V=\{r_1,r_2,\ldots,r_N\}$, $N:=\# V$.
The number of Hamiltonian cycles, denoted by $\mathcal{H}(G)$,
can be written as a $2N$-point function of a lattice field theory in a 
certain limit.
To this end, one introduces O$(n)$ lattice field $\phi_j(r)$
$(r\in V, j=1,\ldots,n)$ with the action
\begin{equation}
S[\vec{\phi}(r)]:=\frac12\sum_{r,r'\in V, j,k=1,\ldots,n} 
         \phi_j(r) (\sqrt{-1}\Delta^{-1}+\epsilon)_{rr'}\delta_{jk}\phi_k(r').
\label{action}
\end{equation}
An infinitesimal parameter $\epsilon \rightarrow+0$ is the convergence factor.
The $N\times N$ matrix $\Delta$ is the adjacency matrix of the graph $G$:
\begin{equation}
  \Delta_{rr'} :=
\left\{
  \begin{array}{ll}
    1 & \text{if $r,r'\in V$ is connected by an $e\in E$},\\
    0 & \text{otherwise}.
  \end{array}
\right.
\end{equation}
It can readily be shown that \cite{OrItDo:hamiltonian}
\begin{equation}
  \mathcal{H}(G)   
= \lim_{n\rightarrow0}\lim_{\epsilon\rightarrow+0}
  \frac1n
  \left|\left\langle \frac{\vec{\phi}^2(r_1)}{2}
  \ldots\frac{\vec{\phi}^2(r_N)}{2}\right\rangle\right|,
\label{field_theory_rep}
\end{equation}
where
\begin{equation}
\langle\cdots\rangle := \frac1Z\int \prod_{r\in V, j=1,\ldots,n} d\phi_j(r)
 \quad (\cdots) \quad e^{-S[\vec{\phi}(r)]}
\end{equation}
with the normalization factor $Z$ to ensure $\langle 1 \rangle=1$.

The proof is diagrammatic. 
Note that the propagator is proportional to the connectivity matrix:
$\langle \phi_j(r)\phi_k(r')\rangle =-\sqrt{-1}\delta_{jk}\Delta_{rr'}$.
When Wick's theorem is applied, each of the surviving diagram after taking 
the $n\rightarrow0$ limit corresponds to a
Hamiltonian cycle of $G$ with an equal weight. 
The limit $n\rightarrow 0$ is taken for discarding disconnected diagrams.

When the graph $G$ is homogeneous, 
there is a mean field saddle point
\begin{equation}
\phi(r)\equiv\phi^{(0)}=\sqrt{-2qi},
\end{equation}
where $q:=\sum_r \Delta_{rr'}$ is the coordination number of $G$.
It gives rise to the estimate \eqref{mean_field_saddle_point_estimate}.

\section{Saddle points  for sublattices} \label{sec:sublattice}
\label{sec:saddle_point}
Given a graph $G=(V,E)$ which is not necessarily homogeneous, 
I look for a saddle point for the integral expression \eqref{field_theory_rep}.

I decompose the set $V$ into classes:
 $V=\sqcup_{a=1}^m V_a$, with $N_a:=\#V_a$.
Let $q_{a}(r)$ 
be the number of edges one of whose endpoints is $r\in V$
and the other belongs to $V_a$.
I only consider such divisions that  $q_{a}(r)=q_{a}(r')$ holds
if $r$ and $r'$ belong to an identical class, say,  $V_b$.
Then it makes sense to define  $q_{ab}:=q_a(r)$.
Because of the relation $q_{ab} N_b = q_{ba} N_a$, $S_{ab}:=q_{ab}
N/N_a$ becomes a symmetric matrix. 

Any graph does have such a decomposition; one is free to take 
$V_a=\{r_a\}, a=1,\ldots,m=N$.
I concentrate, however, on the case where $m$ stays finite
in the limit $N\rightarrow\infty$.
In other words, I focus on $G$ with a sublattice
structure\footnote{
  Note, however, the saddle point method is worth considering as an
  approximation scheme even for the case $V_a=\{r_a\}, m=N$.
  Estimating $\mathcal{H}(G)$ by the saddle point
  method as described 
  below takes only polynomial time in $m(=N)$
  while 
  determining $\mathcal{H}(G)$ exactly for an
  arbitrary graph $G$ is an NP-complete problem \cite{GaJo:NP-completeness}.
}.
Hereafter $V_a$ is referred as a sublattice.

I proceed to evaluating \eqref{field_theory_rep} by the saddle point method. 
Respecting the sublattice structure,
an ansatz for the saddle point configuration
$\vec{\phi}^{(0)}(r)$ for \eqref{field_theory_rep}, 
\begin{equation}
  \vec{\phi}^{(0)}(r) = \sum_{a=1}^m \vec{v}_a \delta_a(r)
\end{equation}
is employed.
Here $\vec{v}_a$ is an $n$-vector variable to be solved, labeled by
$a=1,\ldots,m$.
The membership function $\delta_a(r)$ is defined by 
\begin{equation}
 \delta_a(r):=
 \begin{cases}
   1 & \text{if\ }  r\in V_a,\\
   0 & \text{otherwise}.
 \end{cases}
\end{equation}
In order to calculate $S[\phi^{(0)}]$, one needs to 
know how $\Delta^{-1}$ acts on $\delta_a$.
First, I write the the number of edges connecting $V_a$ and $r\in V$
in two ways as
\begin{equation}
  \sum_{r'\in V} \Delta_{rr'} \delta_a(r') 
= \sum_{b=1}^m \frac{N_a}{N_b}q_{ba} \delta_b(r).
\end{equation}
Rewriting the right hand side in terms of the $m\times
m$ symmetric matrix $S_{ab}$, I have 
\begin{equation}
\frac{N}{N_a}  \sum_{r'\in V} \Delta_{rr'} \delta_a(r') 
=  \sum_{b=1}^m S_{ab} \delta_b(r).
\end{equation}
I apply $\Delta^{-1}S^{-1}$ on both sides to obtain $\Delta^{-1}\delta_a$:
\begin{equation}
  \sum_{b=1}^m (S^{-1})_{ab} \frac{N}{N_b} \delta_b(r)
= \sum_{r'\in V}(\Delta^{-1})_{rr'} \delta_a(r').
\end{equation}
Because of this relation,  $S[\vec{\phi}^{(0)}(r)]$ reduces to 
\begin{equation}
\!\!\!\!
\!\!\!\!
S[\vec{\phi}^{(0)}(r)]=
   - N \ln 2 + \sum_{a=1}^m N_a \ln (\vec{v}_a \cdot \vec{v}_a)
 -\frac{N}{2} \sum_{a,b=1}^m (S^{-1})_{ab}( \vec{v}_a \cdot \vec{v}_b).
 \label{action_sublattice}
\end{equation} 
One can assume that the saddle point is of the form 
$
  \vec{v}_a=(x_a,0,0,\ldots,0)
$
and vary the action \eqref{action_sublattice} with respect to $x_a$.
One obtains $m$ saddle point equations 
\begin{equation}
  \frac{2 N_a}{N} =  \sum_{b=1}^m (S^{-1})_{ab} x_a x_b,
\quad\quad(a=1,\ldots,m)
  \label{def_of_x}
\end{equation}
or equivalently
\begin{equation}
  \sum_{b=1}^m x_b^{-1} q_{ba} x_a^{-1} = \frac12.  
\end{equation}
which are to be solved for $x_a$, $a=1,\ldots,m$. 
This is a set of $m$ quadratic equations with $m$ unknowns.
If the solution to \eqref{def_of_x} is not unique,
the one that minimizes the action \eqref{action_sublattice} should be
selected. 
Putting the solution back into 
\eqref{action_sublattice}, one  obtains an estimate 
\begin{equation}
\mathcal{H}(G) \sim (\omega_{\mathrm{SP}})^N,  \quad\quad
\omega_{\mathrm{SP}}:= \frac{1}{e}
            \prod_{a=1}^m \left(\frac{x_a^2}{2}\right)^{N_a/N}.
\label{estimate_sublattice}
\end{equation}
This generalizes the estimate \eqref{mean_field_saddle_point_estimate}
to the case of inhomogeneous graphs.

\section{Decorated square lattices} \label{sec:examples}
\subsection{Square-diagonal lattice} \label{sec:ex-sd}
\begin{figure}[tb]
  \begin{center}
    \leavevmode
    \includegraphics[scale=1]{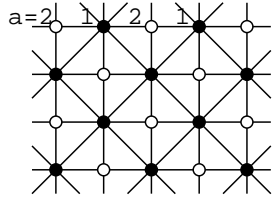}    
  \end{center}
\caption{The square-diagonal (Union Jack) lattice. 
The decomposition of $V$ consists of two
  classes: $V_1=\{\bullet\}$ and $V_2=\{\circ\}$.}
\label{fig:square_diagonal}
\end{figure}

I  apply the  estimate \eqref{estimate_sublattice} to 
the square-diagonal (Union Jack) lattice drawn in
Fig.~\ref{fig:square_diagonal}. 
The decomposition of $V$ consists of two classes 
$V_1=\{\bullet\}$ and $V_2=\{\circ\}$. 
The matrix $q_{ab}$ is given by
\begin{equation}
  q
=
  \begin{pmatrix}
    4 & 4 \\
    4 & 0 \\
  \end{pmatrix}.
\end{equation}
Each class has an equal number of vertices:
$  N_1=N_2=\frac{1}{2} N. $
I assume that the sublattice structure is respected at the boundary.

The field configuration $x_1=8/\delta, x_2=\delta$  becomes a
saddle point in the limit $\delta\rightarrow0$.
Eq. \eqref{estimate_sublattice} yields 
\begin{equation}
  \omega_{\mathrm{SP}}^{\mathrm{SD}}
= \frac{4}{e}
\end{equation}
which is, surprisingly, exactly the same value as the simple  
square lattice. 
It means that the $\mathcal{H}(G)$ of the square-diagonal
lattice grows no faster than that of the simple square lattice
though the former has many additional diagonal edges.
It turns out that it is the case; 
the simple square lattice and the corresponding square-diagonal
lattice have exactly the same number of Hamiltonian cycles, which I
prove rigorously in subsection \ref{sec:rel-sd}. 

\subsection{Square-triangular type lattices}
\begin{figure}[tb]
  \begin{center}
    \leavevmode
    \includegraphics[scale=0.7]{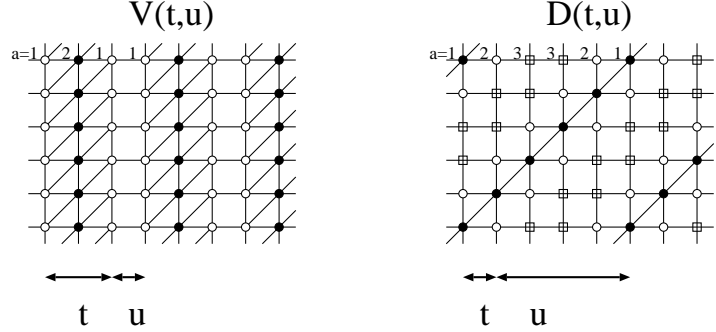}    
  \end{center}
\caption{Square-triangular lattices \protect $V(t=2,u=1)$ and $D(t=1,u=4)$.
The integers $t$ and $u$ are the numbers of plaquettes 
with and without diagonal edges, respectively, consecutive in the
horizontal direction. 
The sublattices are distinguished by the shape of the vertices.
\label{fig:st}
}
\end{figure}

As another example of the use of \eqref{estimate_sublattice}, 
I apply it to a series of inhomogeneous\footnote{
  The lattice $V(1,1)$ turns out to be  a homogeneous graph with 
  the coordination number $q=5$.
} square-triangular lattices $D(t,u)$ and
$V(t,u)$ depicted in Fig.~\ref{fig:st}.
They are square lattices with some of the diagonal edges added.
The matrices $q$ for them are all tridiagonal matrices if one labels 
sublattices appropriately.
Here I show the $m\times m$ matrix $q$ and the fraction $N_a/N$
\ifthenelse{\equal{\version}{jphysa}}{%
  of $V(t,u)$ and $D(t,u)$ with $t=1$ or $u=1$.%
  }{
  of $V(t=1,u=2m-1)$ and $D(t=1,u=2m-3)$, $m=(1),2,3,\ldots$. 
  }

\paragraph{Lattice \protect $V(1,2m-1)$}
\begin{equation}
  q
=
\left(
  \begin{smallmatrix}
     4 & 1 & 0      & 0      & \cdots & 0 \\
     1 & 2 & 1      & 0      &        & \vdots \\
     0 & 1 & 2      & \ddots & \ddots & \vdots \\
     0 & 0 & \ddots & \ddots & 1       & 0 \\
\vdots &   & \ddots & \ddots & 2      & 1 \\
       &         &       &   &        &  \\
     0 & \cdots  & \cdots& 0 & 1      & 3 \\
  \end{smallmatrix}
\right),\quad
\frac{N_a}{N}=\frac{1}{m} \quad(a=1,\ldots,m).
\end{equation}

\ifthenelse{\equal{\version}{condmat}}{%
\paragraph{Lattice \protect $V(1,2m-2)$}
\begin{equation}
  q
=
\left(
  \begin{smallmatrix}
     4 & 1 & 0      & 0      & \cdots & 0 \\
     1 & 2 & 1      & 0      &        & \vdots \\
     0 & 1 & 2      & \ddots & \ddots & \vdots \\
     0 & 0 & \ddots & \ddots & 1       & 0 \\
\vdots &   & \ddots & \ddots & 2      & 2 \\
       &         &       &   &        &  \\
     0 & \cdots  & \cdots& 0 & 1      & 2 \\
  \end{smallmatrix}
\right), \quad
\frac{N_a}{N}=
\begin{cases}
  \frac{1}{2m-1}  & (a=m),\\
  \frac{2}{2m-1}  & (\text{otherwise}).
\end{cases}
\end{equation}

\paragraph{Lattice \protect$V(2m-1,1)$}
\begin{equation}
  q
=
\left(
  \begin{smallmatrix}
     3 & 2 & 0      & 0      & \cdots & 0 \\
     2 & 2 & 2      & 0      &        & \vdots \\
     0 & 2 & 2      & \ddots & \ddots & \vdots \\
     0 & 0 & \ddots & \ddots & 2       & 0 \\
\vdots &   & \ddots & \ddots & 2      & 4 \\
       &         &       &   &        &  \\
     0 & \cdots  & \cdots& 0 & 2      & 2 \\
  \end{smallmatrix}
\right), \quad
\frac{N_a}{N}=\frac{1}{m} \quad (m=1,\ldots,m).
\end{equation}

\paragraph{Lattice \protect $V(2m-2,1)$}
\begin{equation}
  q
=
\left(
  \begin{smallmatrix}
     3 & 2 & 0      & 0      & \cdots & 0 \\
     2 & 2 & 2      & 0      &        & \vdots \\
     0 & 2 & 2      & \ddots & \ddots & \vdots \\
     0 & 0 & \ddots & \ddots & 2       & 0 \\
\vdots &   & \ddots & \ddots & 2      & 2 \\
       &         &       &   &        &  \\
     0 & \cdots  & \cdots& 0 & 2      & 4 \\
  \end{smallmatrix}
\right), \quad
\frac{N_a}{N}=
\begin{cases}
  \frac{1}{2m-1} & (a=m),\\
  \frac{2}{2m-1} & (\text{otherwise}).
\end{cases}
\end{equation}
} 

\paragraph{Lattice \protect $D(1,2m-3)$}
\begin{equation}
  q
=
\left(
  \begin{smallmatrix}
     2 & 2 & 0      & 0      & \cdots & 0 \\
     4 & 0 & 2      & 0      &        & \vdots \\
     0 & 2 & 0      & \ddots & \ddots & \vdots \\
     0 & 0 & \ddots & \ddots & 2      & 0 \\
\vdots &   & \ddots & \ddots & 0      & 4 \\
       &         &       &   &        &  \\
     0 & \cdots  & \cdots &  0 &    2 & 0 \\
  \end{smallmatrix}
\right), \quad
\frac{N_a}{N}=
\begin{cases}
  \frac{1}{2m-2} &(a=1,m),\\
  \frac{2}{2m-2} &(\text{otherwise}).
\end{cases}
\end{equation}

\ifthenelse{\equal{\version}{condmat}}{%
\paragraph{Lattice \protect $D(1,2m-2)$}
\begin{equation}
  q
=
\left(
  \begin{smallmatrix}
     2 & 2 & 0      & 0      & \cdots & 0 \\
     4 & 0 & 2      & 0      &        & \vdots \\
     0 & 2 & 0      & \ddots & \ddots & \vdots \\
     0 & 0 & \ddots & \ddots & \ddots & 0 \\
\vdots &   & \ddots & \ddots & 0      &  2\\
       &         &       &   &        &  \\
     0 & \cdots  & \cdots &  0 &    2 & 2 \\
  \end{smallmatrix}
\right), \quad
\frac{N_a}{N}=
\begin{cases}
  \frac{1}{2m-1} & (a=1),\\
  \frac{2}{2m-1} & (\text{otherwise}).\\
\end{cases}
\end{equation}

\paragraph{Lattice \protect $D(2m-3,1)$ }
\begin{equation}
  q
=
\left(
  \begin{smallmatrix}
     0 & 2 & 0      & 0      & \cdots & 0 \\
     4 & 2 & 2      & 0      &        & \vdots \\
     0 & 2 & 2      & \ddots & \ddots & \vdots \\
     0 & 0 & \ddots & \ddots & 2       & 0 \\
\vdots &   & \ddots & \ddots & 2      & 4 \\
       &         &       &   &        &  \\
     0 & \cdots  & \cdots& 0 & 2      & 2 \\
  \end{smallmatrix}
\right), \quad
\frac{N_a}{N}=
\begin{cases}
  \frac{1}{2m-2} & (a=1,m),\\
  \frac{2}{2m-2} & (\text{otherwise}).\\
\end{cases}
\end{equation}

\paragraph{Lattice $D(2m-2,1)$}
\begin{equation}
  q
=
\left(
  \begin{smallmatrix}
     0 & 2 & 0      & 0      & \cdots & 0 \\
     4 & 2 & 2      & 0      &        & \vdots \\
     0 & 2 & 2      & \ddots & \ddots & \vdots \\
     0 & 0 & \ddots & \ddots & 2       & 0 \\
\vdots &   & \ddots & \ddots & 2      & 2 \\
       &         &       &   &        &  \\
     0 & \cdots  & \cdots &  0 &    2 & 4 \\
  \end{smallmatrix}
\right), \quad
\frac{N_a}{N}=
\begin{cases}
  \frac{1}{2m-1} & (a=1),\\
  \frac{2}{2m-1} & (\text{otherwise}).\\
\end{cases}
\end{equation}
}

\begin{figure}[tb]
  \begin{center}
    \leavevmode
    \includegraphics[scale=1]{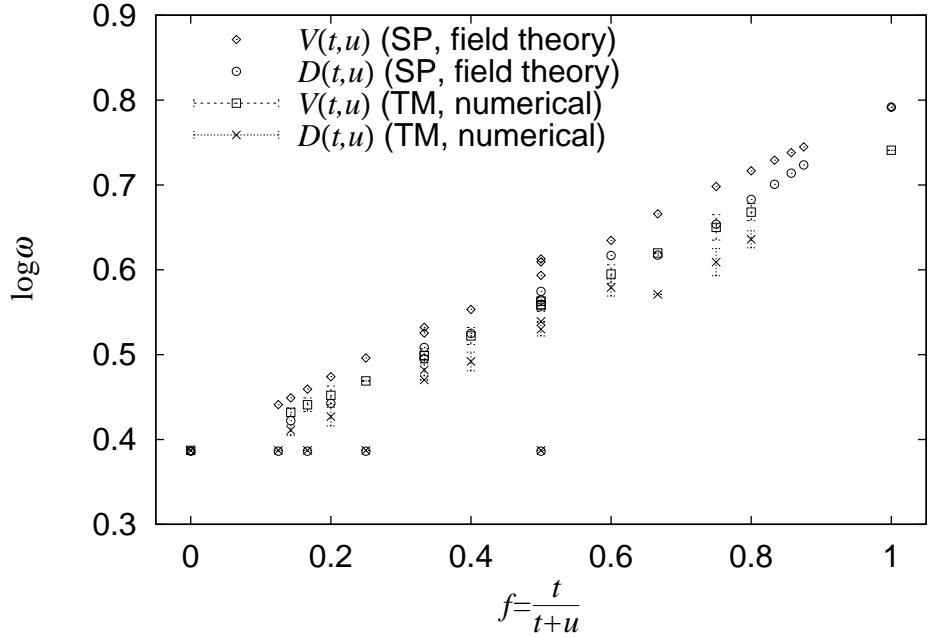}    
  \end{center}
\caption{The estimates for $\omega$ are plotted.
The saddle point approximation 
$\omega_{\mathrm{SP}}$ for $V(t,u)$ ($\diamond$) and $D(t,u)$ ($\circ$) 
with $2t+u\lesssim16$
are compared with the result of transfer matrix calculation
$\omega_\mathrm{TM}$ ( $\square$ and $\times$ with an error bar, respectively).
For details  at $f=1/2,1/3$, see Table~\ref{table:numerical}.
  }
\label{fig:st2}
\end{figure}

For small $m$,  eq. \eqref{def_of_x} can be solved analytically.
For example, $V(1,3)$ turns out to possess the saddle point
\begin{equation}
  \frac{x_1}{2}
= \frac{x_2}{3}
=\sqrt{2+\frac{1}{\sqrt{3}}},
\end{equation}
which implies
\begin{equation}
  \omega_{\mathrm{SP}}^{V(1,3)} = \frac{1+2\sqrt{3}}{e}=1.64225\cdots.
\end{equation}
As $m$ increases, one is forced to solve eq. \eqref{def_of_x}
numerically.
Fig.~\ref{fig:st2} illustrates the dependence of 
$\log\omega_{\mathrm{SP}}$ of $V(t,u)$ ($\diamond$) and $D(t,u)$ ($\circ$) 
on 
$
 f:=t/(t+u).
$
Note that $f=0,1$ correspond to 
the simple square lattice and the triangular lattice, respectively.

For the purpose of comparison, 
I also plot an estimate $\omega_{\mathrm{TM}}$, 
the result of numerical diagonalization of the transfer matrices 
for $V(t,u)$ ($\square$) and $D(t,u)$ ($\times$).
The method to calculate $\omega_{\mathrm{TM}}$ is explained later in 
section \ref{sec:transfer_matrix}. 
Here I only mention that, under plausible assumptions,
$\omega_{\mathrm{TM}}$ agrees with the true value $\omega$
within the precision of the numerical work and the error originating
{from} the finite size effects, though it takes considerable memory and
CPU time to compute it\footnote{
In the present analysis, 
the numerical diagonalization of sparse matrices of dimension
$\lesssim 10^6$ has been carried out.
}. 
On the other hand, calculating $\omega_{\mathrm{SP}}$ is quite easy.

I argue that the estimates $\omega_{\mathrm{SP}}$ capture the 
qualitative feature of $\omega_{\mathrm{TM}}$ though there is a
certain amount of difference in the numerical values.

First of all, 
one notices that $\omega_{\mathrm{SP}}$ for $D(1,2m-3)$ is saturated
just like the square-diagonal lattice:
\begin{equation}
  \omega_{\mathrm{SP}}^{D(1,2m-3)} = \frac{4}{e}  \quad (m=2,3,\ldots).
\end{equation}
This suggests that the diagonal edges contribute almost
nothing to $\mathcal{H}(D(1,2m-3))$. 

In accordance with this fact, 
the numerical estimate $\omega_{\mathrm{TM}}^{D(1,2m-3)}$ also 
stays identical with $\omega_{\mathrm{TM}}$ of the simple square
lattice. 
I will argue in subsection \ref{sec:rel-st}
that $\omega$ for  $D(1,2m-3)$ and
that for the simple square lattice should in fact be identical.

In the field theory, the saddle point for $D(1,2m-3)$ is given by
\begin{equation}
  x_{2a+1}=8/\delta, \quad x_{2a}=\delta \quad(\delta\rightarrow0).
\end{equation}
It is intriguing that one needs the limiting procedure
$\delta\rightarrow0$ whenever one obtains the saturated value
$\omega_{\mathrm{SP}}=4/e$. 

Secondly, apart from the series $D(1,2m-3)$,
the estimate $\omega_{\mathrm{SP}}$ increases with $f$ for both 
$V(t,u)$ and $D(t,u)$ as naturally expected.
The result for $\omega_{\mathrm{TM}}$ confirms this naive expectation.
Moreover, in both analysis, $\omega_{\mathrm{SP/TM}}$ for $D(t,u)$ 
is always slightly lower than $V(t,u)$ for a fixed $(t,u)$.

Lastly, some detailed structures of $\omega_{\mathrm{TM}}$
are reproduced in $\omega_{\mathrm{SP}}$.
For example, there is a violation of simply-increasingness of 
$\omega_{\mathrm{TM}}$ for the pair $D(2,1)$ and $D(3,2)$.
It is present also for $\omega_{\mathrm{SP}}$.

Near $f=1$, there is a considerable difference between 
the numerical values of $\omega_{\mathrm{SP}}$ and $\omega_{\mathrm{TM}}$.
Note, however, this difference is already there for the triangular
lattice, 
for which  the mean field saddle point approximation 
\eqref{mean_field_saddle_point_estimate} has been applied.
Thus, this difference does not imply the fault of 
the extension \eqref{estimate_sublattice}.

\section{Exact relations among $\mathcal{H}(G)$} \label{sec:exact_relations}
I prove some equalities among $\mathcal{H}(G)$ for the square-diagonal
lattice and the square-triangular type lattices.

\subsection{Square-diagonal lattice} \label{sec:rel-sd}
Let $G_{\mathrm{SD}}$ be a square-diagonal lattice of a rectangular 
shape with $L_x\times L_y=N$ vertices.
The periodic boundary condition is imposed across the edges of the rectangle
giving $G_{\mathrm{SD}}$ the toric topology.
I assume that $L_x$ and $L_y$ are even and the square-diagonal lattice 
structure is consistent with the boundary condition.
I define a simple square lattice $G_{\mathrm{SS}}$ as the lattice
obtained by removing all the diagonal edges from $G_{\mathrm{SD}}$. 
It also has $N$ vertices.
I shall prove $\mathcal{H}(G_{\mathrm{SD}})=\mathcal{H}(G_{\mathrm{SS}})$.
In other words, I will show that Hamiltonian cycles on
$G_{\mathrm{SD}}$ passes vertical and horizontal edges but not diagonal edges.

I decompose the set of vertices of $G_{\mathrm{SD}}$ into $V_1$ and
$V_2$ as depicted in  Fig.~\ref{fig:square_diagonal}. Then $N_1=N_2=N/2$.
On a Hamiltonian cycle, there are $N$ vertices and $N$ edges.
Suppose $N^{(d)}$ edges out of $N$ are diagonal ones.
The two ends of a diagonal edge belong to $V_1$.
On the other hand, there are $N-N^{(d)}$ vertical or horizontal edges 
each of which connects $V_1$ and $V_2$.
Thus I have 
\begin{equation}
  N_1 = \left( 2 \times N^{(d)} + 1\times (N-N^{(d)}  ) \right)\times \frac12
\end{equation}
and
\begin{equation}
  N_2 = (N-N^{(d)}) \times \frac12
\end{equation}
which imply $N^{(d)}=0$.

Because $\mathcal{H}(G_{\mathrm{SD}})=\mathcal{H}(G_{\mathrm{SS}})$
for $G_{\mathrm{SD}}$ and $G_{\mathrm{SS}}$ at any finite size,
one can take the limit $N\rightarrow\infty$ to have
$\omega^{\mathrm{SD}}=\omega^{\mathrm{SS}}$.

\subsection{Lattice \protect$D(1,2m-3)$ } \label{sec:rel-st}
Let $G_m$ be a finite $D(1,2m-3)$ lattice of a 
rectangular shape. Again I assume the periodic boundary condition on
both directions, respecting the 
$D(1,2m-3)$ structure at the boundary.
I shall show
$\mathcal{H}(G_m)=\mathcal{H}(G_{\mathrm{SS}})$.

For brevity, I write $G \sqsubset G'$ 
if $V=V'$ and $E\subset E'$,  where $G=(V,E)$ and $G'=(V',E')$.
Apparently, 
$\mathcal{H}(G)\leq\mathcal{H}(G')$
if $G \sqsubset G'$.
We have seen that the inequality 
$\mathcal{H}(G_{\mathrm{SS}})\le \mathcal{H}(G_{\mathrm{SD}})$
for the pair 
$G_{\mathrm{SS}} \sqsubset G_{\mathrm{SD}}$
is actually an equality.

It is  crucial to find that 
$G_{\mathrm{SS}} \sqsubset G_m  \sqsubset G_{\mathrm{SD}}$.
Thus, I obtain 
$\mathcal{H}(G_{\mathrm{SS}})=\mathcal{H}(G_m)$.
This relation, in the limit $N\rightarrow\infty$, again implies 
that 
$\omega^{\mathrm{SS}}=\omega^{D(1,2m-3)}$.

\section{Numerical transfer matrix method} 
\label{sec:transfer_matrix}
In order to estimate  $\omega$ numerically, 
I map the problem of Hamiltonian cycles onto the 
zero fugacity limit of the fully packed loop model.
Then I represent the fully packed model onto a state sum model with a
local weight
in order to construct a row transfer matrix.
The quantity $\omega$ is related to an eigenvalue of the
transfer matrix.

\begin{figure}[tbph]
  \begin{center}
    \leavevmode
    \includegraphics[scale=0.4]{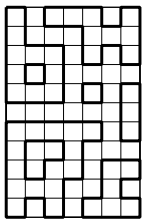}
    \caption{An example of fully packed loop configuration with 
        $\mathcal{N}_\mathrm{L}=8$ on the simple square lattice
        $V(0,1)$ . The loops are
        drawn in thick lines. 
    \label{fig:fplconf}}
  \end{center}
\end{figure}

The partition function of the fully packed model with the
the loop fugacity $n$  is given by 
\begin{equation}
  Z_{\mathrm{FPL}}(n) = \sum_{\text{fully packed loop configuration}}
    n^{\mathcal{N}_\mathrm{L}}.
  \label{fpl-partition}
\end{equation}
The sum in \eqref{fpl-partition} is over all the fully packed loop
configurations (Fig.~\ref{fig:fplconf}), that is non-intersecting
closed loops on the lattice 
such that every vertex is visited by exactly one of the loops.
The number of connected components of loops, denoted by 
$\mathcal{N}_\mathrm{L}$, 
can be greater than one.

\begin{figure}[tb]
  \begin{center}
    \leavevmode
    \includegraphics[scale=0.6]{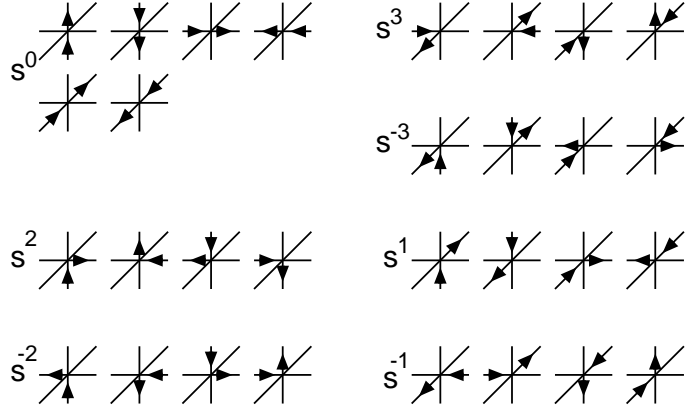}
  \end{center}
  \caption{
    A vertex weight for the state sum model for the fully
    packed loop model. The parameter $s$ is a complex number. 
    Other configurations are prohibited ( weight $W=0$ ).
    \label{fig:weights}
    }
\end{figure}

One can construct an equivalent  state sum model in the following way.
One begins with the triangular lattice $V(1,0)$ with the coordination
number $q=6$. 
The local degrees of freedom $z$ live on each edge $e$.
The three possible values of $z(e)$ is
$\leftarrow$ (a directed edge), 
$\rightarrow$ (an oppositely directed edge ),
and $-$ (a vacant edge).
Let $e_1,\ldots,e_6$ be the edges that share a vertex $r$.
States on $e_1,\ldots,e_6$ interact on $r$.
The local vertex weight is 
\begin{equation}
  W(z(e_1),\ldots,z(e_6))= s^k \quad
  (s\in\mathbb{C})
\label{vertexweight}
\end{equation}
if there is exactly one ingoing arrow and an outgoing one at $r$,
where the integer $k$ is given by
\begin{equation}
  k=\frac{(\text{the angle of the right turn})}{\pi/4}.
  \label{power_of_s}
\end{equation}
as illustrated in Fig.~\ref{fig:weights}. 
Otherwise, $W=0$.

In other lattices $D(t,u)$ and $V(t,u)$, some of the diagonal edges 
in the lattice $V(1,0)$ are missing. 
One extends the above construction to them by just 
interpreting these missing edges as existing edges on which only the
vacant edge state is allowed. 

The partition function of the state sum model is given by 
  \begin{equation}
 Z(s) :=
\sum_{z(e)=\leftarrow, - , \rightarrow  \quad (e\in E)} \quad\quad
\prod_{r\in V}   W(z(e_{j_1}),\ldots,z(e_{j_6})).
  \end{equation}
Edges $e_{j_1},\ldots,e_{j_6}$ are understood to share a vertex $r$.
One sees that the surviving configurations in the state sum model
are nothing but the fully packed loop configurations
 with a direction associated with each loop component. 
Let us inspect a contribution from a loop.
When one walks along the loop in the associated direction to come back
to the original point, one has changed the direction $\pm2\pi$ in
total (if it is a topologically trivial loop).
Therefore, the  product of the vertex weight on the loop is 
$s^{\pm8}$ due to the choice \eqref{power_of_s}.
This is why the partition function of the state sum model agrees with
\eqref{fpl-partition} with  
\begin{equation}
  n=s^8+s^{-8}.
\end{equation}

Now one can think of a row transfer matrix $\mathbf{T}$
of the fully packed loop model. For the lattices $D(t,u)$ and $V(t,u)$, 
one introduces a transfer matrix which maps  a state on a row of 
vertical and horizontal edges
onto that on the upper row in Fig.~\ref{fig:st}.
That is, a state acted by $\mathbf{T}$ is a linear combination of an
arrow configuration on the row $|/|/|/|/\cdots/|/|$. 
For the lattice $D(t,u)$, a one-unit shift in the horizontal
direction is included in $\mathbf{T}$ to take care of the change of
the positions  of diagonal edges. 

It is important to note that the transfer matrix $\mathbf{T}$ commutes
with the operator giving the net number of upward arrows
\begin{equation}
 d= (\#\uparrow) +(\#\nearrow) - (\#\downarrow)- (\#\swarrow).
\end{equation}
Thus  $\mathbf{T}$ is block-diagonalized as 
$\mathbf{T}=\oplus_d \mathbf{T}_{d}$.

\begin{figure}[tb]
  \begin{center}
    \leavevmode
    \includegraphics[scale=0.7]{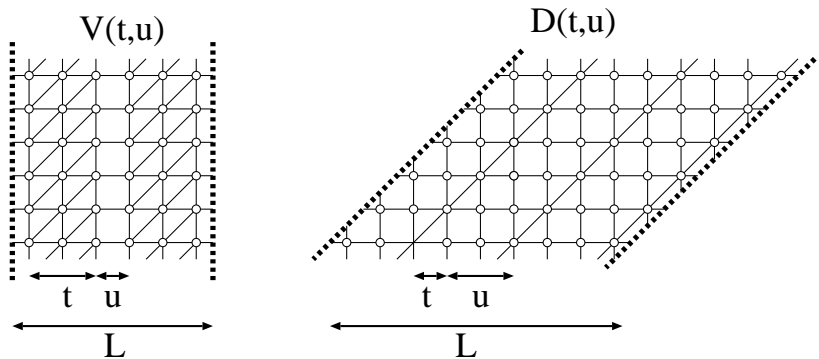}    
  \end{center}
\caption{Square-triangular lattices \protect $V(t=2,u=1)$ and $D(t=1,u=2)$.
The periodic boundary condition in the horizontal direction is
introduced to give the lattice infinite cylinder topology.
The seam is indicated by broken lines.
}
\label{fig:seam}
\end{figure}

One considers an infinite cylinder geometry
which is suitable for numerical transfer matrix calculation.
The horizontal direction is compactified with the period $L$ as indicated in
Fig.~\ref{fig:seam}.
In the $d=0$ sector there can be loops which wind the cylinder once. 
It gives rise to a complication because 
such a loop contributes $s^0+s^0=2$, not $s^8+s^{-8}=n$.
To avoid this, one introduces a seam as in Fig.~\ref{fig:seam}.
One declares that a loop which goes across the seam
{from} the left to the right should gain 
an additional weight $s^{8}$, while the left-going one should gain 
$s^{-8}$.

Hamiltonian cycles are encoded in the $d=0$ sector in the limit
$n\rightarrow0$, 
or  $s\rightarrow \exp[\pm\tfrac{\pi i}{16}]$ on the infinite cylinder.
The condition $d=0$ excludes the configurations which have 
unbalanced strings which travels from an end to the other end of the cylinder.
The limit $n\rightarrow0$ excludes small disconnected loops.
In the following, $n$ is simply set to zero.
The connectivity constant $\omega_{\mathrm{TM}}(L)$ 
for this geometry is given by 
\begin{equation}
  \log\omega_{\mathrm{TM}}(L) = \frac{1}{L} \log |\lambda_0^{0}(L)|,
\end{equation}
where $\lambda^i_d(L)$ is the $i$-th largest (in the absolute value)
eigenvalue of $\mathbf{T}_d(L)$. 
One expects that,  by taking the $L\rightarrow\infty$ limit,
one arrives at the bulk value 
$\omega_{\mathrm{TM}}=\omega_{\mathrm{TM}}(\infty)$.
The values of $\omega_{\mathrm{TM}}$ in  Fig.~\ref{fig:st2} have been
calculated in this way. 

I notice that that the relation 
$\omega_{\mathrm{TM}}^{\mathrm{SS}}
=\omega_{\mathrm{TM}}^{\mathrm{SD}}
=\omega_{\mathrm{TM}}^{D(1,2m-3)}$ holds already at every finite $L$.

I assume that the rotational symmetry is restored 
in the $L\rightarrow\infty$ limit with the sound
velocity $v=1$ 
and that the system is described by a conformal field theory
 \cite{BePoZa:infinite},  
which is the case for the simple square lattice $V(0,1)$.
For other lattice, one observes, at least, the mass gap closes at
$L\rightarrow\infty$.
Then scaling exponents are related to the finite size behavior of
other eigenvalues of the transfer matrix \cite{BlCaNi:finite,Affleck:finite}. 
The central charge $c$ appears as\footnote{For the triangular
  lattice, the restoration of conformal symmetry occurs in the frame 
  where the triangles are regular. Thus the formulae should be modified by
  the factor of $\sqrt{3}/4$.
}
\begin{equation}
 \log \omega_{\mathrm{TM}} (L) 
=  \frac{1}{L} \log |\lambda^0_0(L)| 
= \log \omega_{\mathrm{TM}}(\infty) + \frac{\pi c}{6L^2} + \mathcal{O}(L^{-4}).
\end{equation}
The correlation lengths $\xi_i$ and the scaling dimensions $X_i$ of 
general geometric operators labeled by $i$ are given by \cite{DuSa:exact}
\begin{equation}
  \xi_i^{-1}= \log \frac{|\lambda^{0}_i(L)|}{|\lambda^{0}_0(L)|} =
\frac{2\pi X_i}{L} + \mathcal{O}(L^{-3}).
\end{equation}
In particular, the geometric scaling exponents $X_1$ and $X_2$ 
(corresponding to one and two spanning strings, respectively)
are related to the conformational exponent $\gamma$ in eq.
\eqref{asymptotic} by
\begin{equation}
  \gamma=2(1-X_1)\nu, \quad \frac{1}{\nu} = 2-X_2.
\end{equation}

The result of the finite size scaling analysis is 
shown in Table~\ref{table:numerical}.
\begin{table}[htbp]
  \begin{center}
    \leavevmode
\begin{tabular}{|l|l|l|l|l|l|l|} \hline 
lattice      & $L_{\text{max}}$   &  $c$ & $X_1$ &
$\log\omega_{\mathrm{TM}}$  &
$\log\omega_{\mathrm{SP}}$ 
\\ \hline\hline
$V(0,1)_{L\equiv0\bmod2}$&14&$-1.000(11)$&$-0.04423(81)$&0.387171(1)&0.386294\\
$V(0,1)_{L\equiv1\bmod2}$ &13&$ -2.510(5)$&$-0.1875(10)$&0.387168(2)&0.386294\\
$V(1,0)_{L\equiv0\bmod3}$     & 9&$-2.06(8)$&$-0.196(18)$&0.74088(7)&0.791761\\
$V(1,1)$ & 12   &$-1.98(9)$&$-0.175(17)$&0.563057(18)&0.609438\\
$V(1,2)$ &  12  &$-2.00(4)$&$-0.189(3)$&0.498809(3)&0.532186\\
$D(1,2)$ &  12  &$-1.98(10)$&$-0.186(25)$&0.470379(3)&0.495166\\
$V(2,1)$ &  9    &$-1.93(9)$&$-0.183(19)$&0.619823(4)&0.665920\\
$D(2,1)$ &  9    &$-1.98(3)$&$-0.189(14)$&0.571025(1)&0.617343\\
$V(1,3)$ &  12   &$-1.90(29)$&$-0.176(3)$&  0.46905(4)&0.496068\\
$V(2,2)$ &  12  &$-1.89(18) $&$-0.216(10)$&0.55761(4)&0.612848\\
$D(2,2)$ &  12  &$-1.90(14) $&$-0.203(10)$&0.53917(3)&0.574521\\
$V(3,1)$ &  8      &$-1.87(18)$&$-0.174(10)$&0.650(15)&0.698199\\
$D(3,1)$ &  8      &$-2.01(18)$&$-0.187(10)$&0.609(16)&0.653694\\
$V(1,4)$ &  10     &$-2.09(11)$&$-0.187(4)$&0.452(11)&0.474064\\
$D(1,4)$ &  10     &$-2.10(11)$&$-0.177(4)$&0.427(11)&0.442671\\
$V(2,3)$ &  10     &$-1.94(11)$&$-0.170(3)$&0.522(10)&0.553356\\
$D(2,3)$ &  10     &$-2.08(11)$&$-0.189(3)$&0.492(11)&0.524924\\
$V(3,2)$ &  10     &$-2.02(11)$&$-0.189(3)$&0.595(11)&0.634758\\
$D(3,2)$ &  10     &$-1.99(11)$&$-0.185(3)$&0.579(10)&0.616846\\
$V(4,1)$ &  10     &$-1.95(11)$&$-0.184(3)$&0.668(10)&0.716732\\
$D(4,1)$ &  10     &$-1.92(11)$&$-0.180(3)$&0.636(10)&0.683031\\
$V(1,5)$ &  12     &$-2.24(8) $&$-0.213(3)$&0.441(8)&0.459443\\
$V(2,4)$ &  12     &$-2.19(8) $&$-0.210(3)$&0.499(8)&0.525490\\
$D(2,4)$ &  12     &$-2.07(8) $&$-0.192(3)$&0.482(7)&0.508479\\
$V(3,3)$ &  12     &$-2.04(8) $&\multicolumn{1}{|c|}{---}&0.559(8)& 0.593499\\
$D(3,3)$ &  12     & -2.00(8) &\multicolumn{1}{|c|}{---}& 0.530(8)& 0.564561\\
$V(1,6)$ &  14     &$-2.02(6) $&\multicolumn{1}{|c|}{---}&0.432(5)&0.459443\\
$D(1,6)$ &  14     &$-2.10(6) $&\multicolumn{1}{|c|}{---}&0.411(6)&0.386294\\
\hline
\end{tabular}  
    \caption{Exponents and the entropy per vertex estimated by the
      numerical transfer 
matrix method. 
Lattices $D(1,2m-3)$ are suppressed because they share the properties
with $V(0,1)$.
The integer $L_{\mathrm{max}}$ is the largest $L$, 
the size of the lattice in the horizontal direction, which has been  examined.
The exponent $X_2$ is also measured and is found to be zero for all
the lattices.
    \label{table:numerical}}
  \end{center}
\end{table}
I have treated even and odd $L$ separately 
for $V(0,1)$ because there is an oscillatory behavior of period 2 due
to the twist-like operator insertion \cite{BaBlNiYu:packedloop}.
I also see a period-3 oscillation for $V(1,0)$.
Presence of such oscillations is not assumed for analysis of other lattices.
I find that the exponent $X_2$ is always zero at finite $L$ and is supposed to
be so in the limit $L\rightarrow\infty$.

The result in Table~\ref{table:numerical} 
suggests that, except for $V(0,1)(=D(0,1))$ and $D(1,2m-3)$,
the system lies in the same universality class as
the dense phase of O($n$) loop model at $n=0$ whose exponents are
\begin{equation}
\begin{split}
c=-2,\quad
 X_1=-\frac{3}{16}=-0.1875, \quad X_2=0,\\
\nu=\frac{1}{2}, \quad \gamma=\frac{19}{16}=1.1875.  
\end{split}
\end{equation}

This fact can be understood well by regarding  the fully packed loop
model as a special case of the O($n$) loop model whose partition
function is given by 
\begin{equation}
  Z_{\mathrm{loop}}(n,x^{-1}) = \sum_{\text{loop configuration}}
    x^{\mathcal{N}_\mathrm{V}-N} n^{\mathcal{N}_\mathrm{L}},
  \label{loop-partition}
\end{equation}
where the summation is over all the non-intersecting loop configurations
not necessarily fully packed. 
Thus, the number $\mathcal{N}_\mathrm{V}$ of
vertices visited by a loop can be different from $N$.
In the two parameter space $(n,x^{-1})$, 
the dense phase emerges in the region where the O($n$) temperature
$|x^{-1}|$ is small enough
while the fully packed loop model is reproduced by setting $x^{-1}$ to
exactly zero. 

For the simple square lattice and the hexagonal lattice, the line
$x^{-1}=0$ is an unstable critical line where a new universality emerges.
This is due to the symmetry $x^{-1}\leftrightarrow -x^{-1}$.
Actually, this symmetry holds because any loops on the simple square
lattice visit even number of vertices.

The result in Table~\ref{table:numerical} means that $x^{-1}=0$ is not
special for other lattices. This is because they admit loops which
visit odd number of vertices \cite{BaBlNiYu:packedloop}. 

\section{Summary} \label{sec:summary}
I have estimated the number Hamiltonian cycles on inhomogeneous
graphs analytically and numerically.
To estimate it analytically, I have employed the field theoretic representation
and have applied the saddle point method. A formula for $\omega$ for graphs
with sublattice structures has been obtained.
The numerical estimation is based on the diagonalized the transfer
matrix of the fully packed model  on the infinite cylinder geometry.

The former result is simple and is believed to capture the physics of
Hamiltonian cycles.
The latter is accurate and provides with the scaling exponents though
it  spends lots of computer time. The results agree each other
qualitatively. 
It is confirmed that the success of the estimate
$\omega_\mathrm{SP}=3/e,4/e$ and $6/e$ by the former method
was not accidental; it now works fine 
for a family of lattices yielding various values of $\omega_\mathrm{SP}$.
Moreover, the former method successfully predicts the
relation $\omega^{\mathrm{SS}}=\omega^{\mathrm{SD}}=\omega^{D(1,2m-3)}$.

The relation 
$\omega_{\mathrm{SP}}^{G}\leq \omega_{\mathrm{SP}}^{G'}$ 
holds for all the pairs $G\sqsubset G'$ I have examined so far.
It may be proved that this holds generally.

The latter method is useful in calculating the exact value of
$\mathcal{H}(G)$ of $G$ with the planar or cylinder topology.
There seems to be, however, no simple  way to apply this method 
to the lattices with the torus and the higher-genus topology, and
three-dimensional lattices.
This is due to the presence of numerous topological sectors of
self-avoiding loops on the lattice \cite{Higuchi:highergenushamiltonian}.
In contrast,  the field theoretic approach was able to predict a 
boundary condition dependence of $\mathcal{H}(G)$ for a family of
toric lattices \cite{cond-mat/9711152}.
Therefore two approaches are considered complementary each other.

\appendix
\section*{Acknowledgments}
Useful conversations with Shinobu Hikami, Kei Minakuchi, Masao Ogata
and Junji Suzuki  
are gratefully acknowledged.
I thank Toshihiro Iwai, Yoshio Uwano, and Shogo Tanimura 
for discussions and hospitality at Kyoto University, where a part of
this work was done.
This work was supported by the Ministry of Education, Science
and Culture under Grant 08454106 and 10740108 
and by Japan Science and Technology Corporation under CREST.  \medskip

\ifthenelse{\equal{\version}{condmat}}{
  \bibliographystyle{apply}
  }{
  \bibliographystyle{physlett}
  }

\bibliography{mrabbrev,shrtjour,polymer,textbook,cft,mypubl}
\end{document}